\documentclass{emulateapj}  
\renewcommand{\b}[1]{\mathbf{#1}} 
\newcommand{\unit}[1]{\nobreak{\mathrm{\;#1}}} 
\newcommand{\msc}[1]{\nobreak{\textrm{\tiny{#1}}}} 
\newcommand{\ud}{{\rm d}} 

\newcommand{\bJ}{\mathbf{J}}
\newcommand{\hb}[1]{\hat{\mathbf{#1}}}
\newcommand{\ba}{\hat{\mathbf{a}}_1}

\newcommand{\cross}{\times}
\newcommand{\be}{\begin{equation}}
\newcommand{\ee}{\end{equation}}
\newcommand{\eq}[1]{eq.~(\ref{eq:#1})}
\newcommand{\fig}[1]{Fig.~\ref{fig:#1}}
\newcommand{\temp}[1]{T_{\rm #1}}
\newcommand{\Tia}{T_{\rm ia}}
\newcommand{\um}{\,\mu{\rm m}}
\newcommand{\NC}{N_{\msc{C}}}
\newcommand{\pan}[1]{\,\textit{#1}}

\begin{document}
\title{Polarized Infrared Emission by Polycyclic Aromatic Hydrocarbons resulting from Anisotropic Illumination}
\author{Lorenzo Sironi and  Bruce T. Draine}
\affil{Princeton University Observatory, Peyton Hall, Princeton, NJ 08544-1001}
\email{lsironi@astro.princeton.edu;\\ draine@astro.princeton.edu}

\begin{abstract}
We study the polarized infrared emission by Polycyclic Aromatic
Hydrocarbons (PAHs), when anisotropically illuminated by
UV photons. PAH molecules are modeled as planar disks with in-plane and
out-of-plane vibrational dipoles. As first pointed out by Leger~(1988), infrared emission features resulting from in-plane and
out-of-plane modes should have orthogonal polarization directions. We
show analytically how the degree of polarization depends on the
viewing geometry and the molecule's internal alignment between
principal axis of inertia and angular momentum, which gets worse
after photon absorption. Longer wavelength features, emitted after
better internal alignment is recovered, should be more
strongly polarized. The degree of polarization for uni-directional
illumination (e.g., by a star) is larger than for diffuse illumination
(e.g., by a disk galaxy), all else being equal. For PAHs in the Cold
Neutral Medium, the predicted polarization is probably too small to distinguish
from the contribution of linear dichroism by aligned foreground
dust. 
The level of polarization predicted for PAH emission from the Orion Bar
is only $\approx0.06\%$ at 3.3$\,\mu$m;
Sellgren et al.~(1988) report a much larger value, $0.86\pm0.28\,\%$,
which suggests that the smallest PAHs may have moderately
suprathermal rotation rates. Future observations of (or upper limits
on) the degree of polarization for the Orion Bar or for dust above
edge-on galaxies (e.g., NGC~891 or M82) may constrain the internal
alignment of emitting PAHs, thus providing clues to their rotational
dynamics. 
\end{abstract}
\keywords{ISM: dust, extinction --- ISM: general --- infrared: galaxies}

\section{Introduction}
The strong infrared emission features at 3.3, 6.2, 7.7, 8.6, 11.3 and 12.7$\,\mu$m
have been attributed to vibrational modes in planar Polycyclic
Aromatic Hydrocarbons (PAHs) \citep[e.g.,][]{leger&puget_84,
allamandola_85}. 
Additional strong features at 16.4 and $\sim17\,\mu{\rm m}$ 
\citep[e.g.,][]{Smith+Draine+Dale+etal_2007}
have also been attributed to PAHs, although the identification is less
certain.
\citet{leger_88} noted that a planar PAH
molecule may emit partially polarized light if anisotropically
illuminated by a source of UV photons. The basic reasons are the
following: \textit{i}) UV absorption is favored if the molecule faces
the illuminating source; \textit{ii}) spinning of the molecule around
its angular momentum preserves some memory of the illumination
direction; \textit{iii}) the vibrational dipoles responsible for the
IR emission features oscillate either perpendicular or parallel to the
molecular plane. 
In particular, the C-H stretching mode (3.3$\,\mu$m) and the 
in-plane C-H bending mode (8.6$\,\mu$m)
oscillate parallel to the grain plane,
whereas the out-of-plane C-H bending mode (11.3 and 
12.7$\,\mu$m) oscillates perpendicular to the molecular plane. 
The strong emission features at $6.2$ and $7.7\um$ are
believed to arise from in-plane C-C stretching and bending modes.
In-plane modes
(3.3, 6.2, 7.7, 8.6$\,\um$) and out-of-plane modes (11.3, 12.7$\,\mu$m)
should exhibit orthogonal polarization angles \citep{leger_88}, and
their electric field vector should be respectively perpendicular and parallel
to the plane-of-sky projection of the illumination direction.

\citet{sellgren_88} have searched for linear polarization of the 3.3
 and 11.3 $\mu$m emission features in a variety of astronomical
sources where PAH emission is observed offset from the
illuminating source. Their upper limits on the
degree of polarization are of the order a few percent.
At one position on the
Orion Bar they measure a linear polarization of $0.86\pm0.28\,\%$
in the 3.3$\,\mu$m feature, with the polarization angle
consistent with being orthogonal to the line between the nebula and
the star, as predicted by \citet{leger_88}.
However, as we show below, the polarization 
they report for the $3.3\,\mu$m feature is much larger than expected.

In this work we present analytic formulae for the degree of
polarization of the PAH emission features, when the emitting grains
are anisotropically illuminated. We model PAH molecules as planar
disks with in-plane and out-of-plane vibrational dipoles. We extend the
calculations by \citet{leger_88} to allow for an arbitrary degree of
disalignment between the molecule's principal axis of inertia $\ba$
(perpendicular to the molecular plane) and its angular momentum $\bJ$. We
discuss both the case of a point-like illuminating source, which may
be applied to reflection nebulae like the Orion Bar, and of an
extended source (e.g., a disk galaxy), which may be relevant
for dust above  NGC~891 or M82.
The level of polarization is sensitive to the angle between
the line of sight and the illumination direction, and to the degree of
alignment between $\ba$ and $\bJ$.
Measurements of the degree of polarization 
can therefore provide insight into the rotational dynamics of PAHs.

This work is organized as follows: 
in \S \ref{sec:model} we describe our model for
UV absorption and polarized IR emission by PAH molecules, commenting
on uncertainties regarding the alignment of $\ba$ with $\bJ$; 
in \S \ref{sec:results}
we present our results, both for a star-like illuminating source and
for an extended galactic disk. 
The reader interested primarily in the observational implications of our work may wish to skip \S \ref{sec:model} and \S \ref{sec:results} and proceed directly to  \S \ref{sec:concl}, where we summarize our findings and discuss how future polarization measurements may constrain the geometrical and rotational properties of PAHs.

\section{Model for polarized emission from PAH\tiny{s}}\label{sec:model}
As discussed by \citet{leger_88}, planar PAH molecules may emit
partially polarized light as a result of anisotropic illumination by a
source of UV photons. UV absorption is favored if the molecular plane is perpendicular to the illumination direction. Following UV
absorption, in-plane and out-of-plane vibrational modes are
excited, producing the observed IR emission features.

The grain angular momentum $\bJ$ stays approximately constant during
the whole process of UV absorption and IR emission \citep{leger_88}:
first, the angular momentum contributed by the absorbed UV photon or
removed via vibrational IR emission or rotational radio emission is
small compared to the mean  angular momentum of interstellar PAHs; secondly, collisions of the emitting grain with interstellar
 atoms or ions hardly occur during the few seconds of IR emission;
finally, Larmor precession of $\bJ$ around the interstellar magnetic
field takes much longer than the IR emission burst
\citep{rouan_92}. 
With $\bJ$ conserved,
some memory of the source direction is retained and
the IR emission
bands will be partially polarized.

\begin{figure*}[htbp]
\centering 
\includegraphics[width=\textwidth]{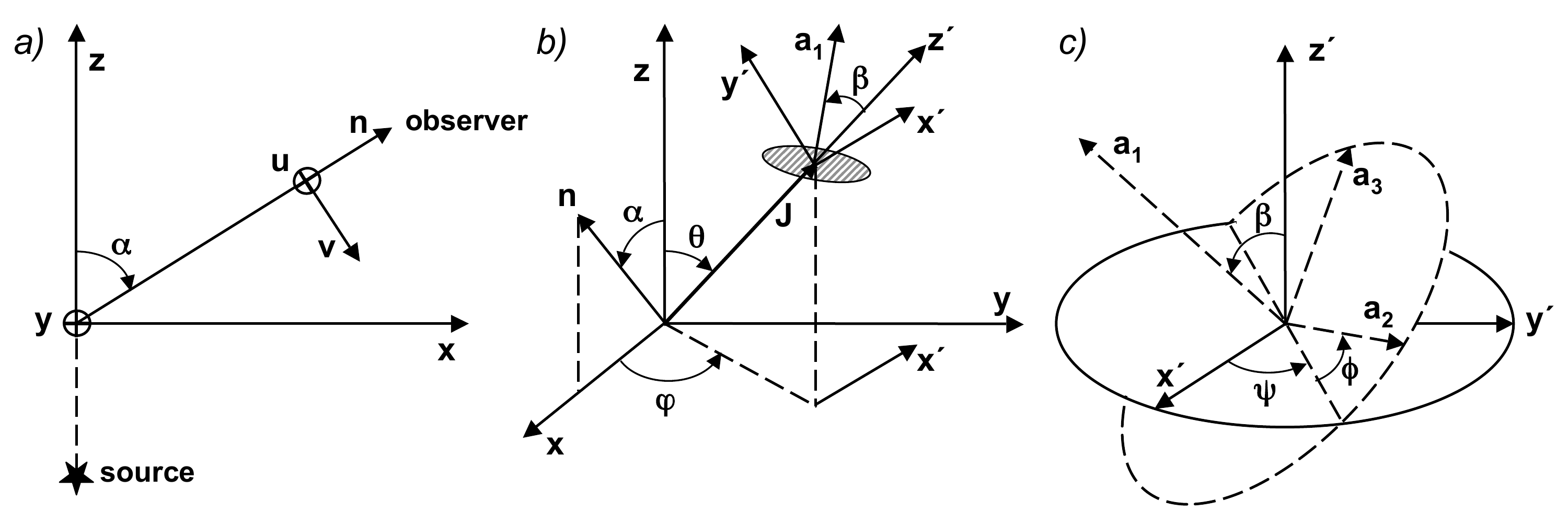}
\caption{Geometry of the star-molecule-observer system, as presented by \citet{leger_88}. In \textit{b}), the molecule (dashed disk) has been moved far from the center of $(\hat{\b{x}},\hat{\b{y}},\hat{\b{z}})$ for clarity. See \S \ref{sec:geo} for details.}
\label{fig:geom} 
\end{figure*}

\subsection{Illumination geometry}\label{sec:geo}
We adopt the system of coordinates used by \citet{leger_88} to specify
the illumination geometry and the orientation of the emitting molecule
(\fig{geom}). The fixed coordinate system
$(\hat{\b{x}},\hat{\b{y}},\hat{\b{z}})$ is centered on the emitting
grain; the polar axis $\hat{\b{z}}$ is along the
illumination direction and $\hat{\b{x}}$ is in the plane defined by
$\hat{\b{z}}$ and the direction $\hat{\b{n}}$ from the molecule to the
observer (\fig{geom}\pan{a}). In this frame, the grain angular momentum $\bJ$ has
spherical angles $\theta$ and $\varphi$ (\fig{geom}\pan{b}). 
In order to
describe the position of the molecule with respect to $\bJ$, we define
a frame $(\hat{\b{x}}',\hat{\b{y}}',\hat{\b{z}}')$ centered on the
grain with $\hat{\b{z}}'$ along $\bJ$ and $\hat{\b{x}}'$ perpendicular
to the plane $(\hat{\b{z}},\hat{\b{z}}')$, as shown in \fig{geom}\pan{b}. 
We choose the grain axes of inertia
$(\hat{\b{a}}_1,\hat{\b{a}}_2,\hat{\b{a}}_3)$ so that $\ba$ is the
axis of largest moment of inertia (i.e., the principal axis); 
for a disk molecule, $\ba$ is perpendicular to the
plane of the molecule, 
whereas $\hat{\b{a}}_2$ and $\hat{\b{a}}_3$ are in the plane. 
Their position in the frame
$(\hat{\b{x}}',\hat{\b{y}}',\hat{\b{z}}')$ is described with Euler's
angles (\fig{geom}\pan{c}): $\beta$, nutation angle, between $\hat{\b{z}}'$ (or
$\bJ$) and $\ba$; $\psi$, precession angle, between $\hat{\b{x}}'$ and
the line of nodes, i.e., the intersection of the plane
$(\hat{\b{x}}',\hat{\b{y}}'$) with the molecular plane
($\hat{\b{a}}_2,\hat{\b{a}}_3$); $\phi$, angle of proper rotation,
between the line of nodes and $\hat{\b{a}}_2$. 
The classical motion of
a rigid axisymmetric grain is a combined rotation around its symmetry
axis $\ba$ and a precession of this axis around the angular momentum
$\bJ$ with constant $\beta.$\footnote{For a perfectly symmetric grain,
the moments of inertia $I_2$ and $I_3$ corresponding to
$\hat{\b{a}}_2$ and $\hat{\b{a}}_3$ are equal and no nutation
occurs. } To describe the polarization of the emitted radiation, we
define $\alpha$, the angle between the illumination direction
$\hat{\b{z}}$ and the observer direction $\hat{\b{n}}$, and the
polarization directions $\hat{\b{u}}$ and $\hat{\b{v}}$, respectively
parallel and perpendicular to $\hat{\b{y}}$ in the plane of the sky
(\fig{geom}\pan{a}).

\subsection{Cross section for absorption of starlight} \label{sec:abs}
The absorption cross section for incident light with electric field
$\b{E}$ is proportional to \citep{leger_88}
\begin{equation}
A\propto\sum_{i,j}|\langle j|\b{E}\cdot\b{d}|i\rangle|^2 ~~~,
\end{equation}
where the summation is over molecular states $|i\rangle$ and
$|j\rangle$ and $\b{d}$ is the electric dipole moment operator.
The $\pi-\pi^\ast$ electronic transitions responsible for UV
absorption in PAHs have $\langle j|\b{d}|i\rangle$ 
only in the molecular plane. For a disk
molecule, rapid spinning around its principal axis $\ba$ results in
averaging over the angle of proper rotation
$\phi$. Thus, the grain absorption cross section may be written, for
incident unpolarized light,
\begin{equation}
A\propto(1+\cos^2\Theta) ~~~,
\end{equation}
where $\Theta$ is the angle between the normal to the grain plane
(i.e., the principal axis $\ba$, for a disk molecule) and the direction of
propagation $\hb{k}$ of the absorbed photon. In other words, when a
planar PAH faces an unpolarized source, its UV absorption cross
section is twice that when it is edge-on, because both components of
the illuminating electric field can be absorbed in the first case and
only one in the second.

The angle $\Theta$ depends on the instantaneous orientation of the
grain with respect to the illumination direction. However, since the precession period of
$\ba$ around $\bJ$ is much shorter than the time between absorption
and emission, we can average over the precession motion (and the precession angle $\psi$) for the absorption and the emission process independently. For a point-like
illuminating source, all incoming rays have $\hb{k}=\hb{z}$ (see \fig{geom}\pan{a}). The
corresponding $\psi$-averaged absorption cross section, for fixed $\bJ$ and a fixed
angle $\beta$ between $\ba$ and $\bJ$, is
\begin{eqnarray}\label{eq:absstar}
\bar A_\star(\theta,\beta)&=&1+\cos^2\theta\cos^2\beta+\frac{1}{2}\sin^2\theta\sin^2\beta\nonumber\\&\equiv&1+C(\theta,\beta) ~~~.
\end{eqnarray}

If the illuminating source is an extended disk galaxy, the angle
$\Theta$ also depends on the spherical angles $\theta'$ and $\varphi'$
of the ray direction $\hb{k}$ in the fixed
$(\hat{\b{x}},\hat{\b{y}},\hat{\b{z}})$ frame. We assume that the
emitting molecules are above the galactic center; $\hat{\b z}$ is then
the direction from the center of the galactic disk to the grains. For
an axisymmetric surface-brightness profile, the absorption coefficient
for incoming rays with polar angle $\theta'$ can be averaged over the
azimuthal illumination angle $\varphi'$:
\begin{equation}\label{eq:absgal}
\bar A_{\rm{gal}}(\theta,\beta, \theta')=1+\cos^2\theta' C(\theta,\beta)+\sin^2\theta'S(\theta,\beta) ~~~,
\end{equation}
where we have also averaged over the precession angle $\psi$. In the
previous expression, $C(\theta,\beta)$ is the same as in
eq.~(\ref{eq:absstar}) and we define
\begin{equation}
S(\theta,\beta)\equiv\frac{1}{4}(1+\cos^2\theta)\sin^2\beta+\frac{1}{2}\sin^2\theta\cos^2\beta ~~~.
\end{equation}
For the sake of simplicity, in the following we assume a uniform-brightness disk galaxy; for a generic axisymmetric brightness profile $B(\theta')$, the absorption cross section in eq.~(\ref{eq:absgal}) should be convolved with $B(\theta')$. As a special case, we observe that if $\theta'=0$ for all incoming rays, i.e. $\hb{k}=\hb{z}$, we recover the absorption cross section in eq.~(\ref{eq:absstar}) for a point-like illuminating source.

\subsection{Cross section for polarized emission}\label{sec:em}
The cross section for PAH emission polarized along a direction
$\hb{w}$ is proportional to \citep{leger_88}
\begin{equation}\label{eq:emission}
F_w\propto\sum_{j,k}|\langle k|\hb{w}\cdot\b{d}|j\rangle|^2 ~~~,
\end{equation}
where $|j\rangle$ and $|k\rangle$ are vibrational states of the emitting molecule. We shall call $F_w^\parallel$ and $F_w^\perp$ the relative emission cross sections for in-plane and out-of-plane modes respectively. 
For out-of-plane modes, the only non-vanishing terms in eq.~(\ref{eq:emission})  involve the component of $\b{d}$ perpendicular to the molecule plane, and  
\begin{equation}
F_w^\perp=(\ba\cdot \hb{w})^2 ~~~.
\end{equation}
For in-plane modes, the matrix elements in eq.~(\ref{eq:emission}) are non-zero only for the components of $\b{d}$ in the molecule plane; for a rapidly-spinning disk molecule, rotational invariance with respect to the angle of proper rotation $\phi$ yields
\begin{equation}\label{eq:em0perp}
F_w^\parallel=1-(\ba\cdot \hb{w})^2=1-F_w^\perp ~~~.
\end{equation}
By averaging over the precession angle $\psi$ for the emission process independently from the absorption (as discussed in \S \ref{sec:abs}), the emission cross sections for out-of-plane modes with polarization along $\hb{u}$ and $\hb{v}$ are
\begin{eqnarray}\label{eq:em1}
\bar F_u^\perp(\theta,\varphi,\beta, \alpha)&=&\frac{1}{2}(\cos^2\varphi+\cos^2\theta\sin^2\varphi)\sin^2\beta\nonumber\\ & &+\sin^2\theta\sin^2\varphi\cos^2\beta ~~~,
\\
\bar F_v^\perp(\theta,\varphi,\beta,\alpha)&=&\cos^2\alpha\left[\frac{1}{2}(\sin^2\varphi+\cos^2\theta\cos^2\varphi)\sin^2\beta\right]\nonumber\\ & &+\cos^2\!\alpha\sin^2\!\theta\cos^2\!\varphi\cos^2\!\beta\!+\!\sin^2\!\alpha\,C(\theta,\beta)\nonumber\\
&&+\frac{1}{4}\sin2\alpha\sin2\theta\cos\varphi(1-3\cos^2\beta) ~,~~
\label{eq:em2}
\end{eqnarray}
where $C(\theta,\beta)$ has been defined in eq.~(\ref{eq:absstar}). We
have made use of the fact that the orientation of the angular momentum
$\bJ$ is constant during the IR emission burst, as discussed at the
beginning of \S \ref{sec:model}.

The corresponding emission cross sections for in-plane modes with polarization along $\hb{u}$ and $\hb{v}$ can be computed from their out-of-plane counterparts by using eq.~(\ref{eq:em0perp}):
\begin{eqnarray}\label{eq:em3}
\bar F_u^\parallel(\theta,\varphi,\beta,\alpha)&=&1-\bar F_u^\perp(\theta,\varphi,\beta,\alpha) ~~~,
\\
\bar F_v^\parallel(\theta,\varphi,\beta,\alpha)&=&1-\bar F_v^\perp(\theta,\varphi,\beta,\alpha) ~~~.
\label{eq:em4}
\end{eqnarray}
By averaging over the azimuthal angle $\varphi$ of the angular momentum distribution, we simplify eqs.~(\ref{eq:em1})-(\ref{eq:em2}) and obtain
\begin{eqnarray}
\bar F_u^\perp(\theta,\beta,\alpha)&=&S(\theta,\beta) ~~~,
\\
\bar F_v^\perp(\theta,\beta,\alpha)&=&\cos^2\alpha \,S(\theta,\beta)+\sin^2\alpha\,C(\theta,\beta) ~~~,
\end{eqnarray}
and the corresponding $\varphi$-averaged cross sections for in-plane
modes may then be derived from eq.~(\ref{eq:em0perp}). In the
following, we assume that the angular momentum $\bJ$ is randomly
oriented in space, but in principle eqs.~(\ref{eq:em1})-(\ref{eq:em2})
and eqs.~(\ref{eq:em3})-(\ref{eq:em4}) could be convolved with any
angular distribution for $\bJ$, e.g., if the emitting molecules are
partially aligned by the interstellar magnetic field.

 \subsection{Model for the alignment of $\ba$ with $\bJ$}\label{sec:align}
The probability that a symmetric grain rotates with angular
momentum $J\equiv|\bJ|$ and nutation angle $\beta$ (between $\ba$ and
$\bJ$) may be written \citep[e.g.,][]{lazarian&draine_97}
\be\label{eq:general}
 \ud P(J,\beta)\propto
f(J)\exp\left(-\frac{E_{\rm {rot}}(J,\beta)}{k_{_{\rm{B}}}T_{\rm
ia}}\right)\sin\beta\,\ud\beta\, \ud J ~~~,
\ee 
where $f(J)\ud J$ is the probability 
that the angular momentum $\in [J,J+\ud J]$,
$E_{\rm rot}$ is the grain rotational energy and $\Tia$ is the
``internal alignment'' temperature which parametrizes the degree of
alignment between $\ba$ and $\bJ$. In the classical approximation, the
rotational energy of a symmetric oblate grain is
\begin{equation}\label{eq:enrot}
E_{\rm{rot}}(J,\beta)=\frac{J^2}{2 I_1}\left[1+\left(\frac{I_1}{I_2}-1\right)\sin^2\beta\right] ~~~,
\end{equation}
where $I_1$ is the largest moment of inertia, corresponding to $\ba$,
and $I_2=I_3<I_1$; for a planar symmetric molecule,
$I_2=I_3=I_1/2$. For the sake of simplicity, we assume that all the
emitting molecules have the same moments of inertia.

The absorption (eqs.~(\ref{eq:absstar})-(\ref{eq:absgal})) and
emission (eqs.~(\ref{eq:em1})-(\ref{eq:em4})) cross sections do not
depend on $J$ but only on the angle $\beta$ between $\ba$ and $\bJ$.
Therefore, we may integrate eq.~(\ref{eq:general}) with respect
to $J$, provided $f(J)$ is known, and the resulting probability
distribution will be used to describe the degree of
alignment between $\ba$ and $\bJ$ when we compute the PAH polarized
emission in \S \ref{sec:results}. In the simple case $f(J)\propto\delta(J-\bar{J})$
(or, in general, if $f(J)$ is strongly peaked at $\bar{J}$), the
probability distribution in eq.~(\ref{eq:general}) reduces to
\begin{equation}\label{eq:align}
\mathrm{d} P_\gamma(\beta)=\frac{\sqrt{\gamma}\exp(-\gamma\sin^2\beta)\sin\beta\,\rm{d}\beta}{\sqrt{\pi}\,e^{-\gamma}\,\mathrm{erfi}(\sqrt{\gamma})}
~~~,
\end{equation}
where $\mathrm{erfi}(z)\equiv(2/\sqrt{\pi})\int_0^z\exp(t^2)\,\ud t$
is the imaginary error function. We have defined  a dimensionless ``internal alignment'' coefficient
\begin{equation}\label{eq:gam_def}
\gamma\equiv\frac{\bar{J}^2}{2\,I_1 k_{_{\rm{B}}}\Tia}\left(\frac{I_1}{I_2}-1\right)\equiv\frac{T_{\rm rot}}{\Tia}\left(\frac{I_1}{I_2}-1\right)
~~~,
\end{equation}
where $T_{\rm rot}\equiv \bar{J}^2/2\,I_1 k_{_{\rm{B}}}$.
When $\gamma\rightarrow\infty$, the molecule principal axis 
tends to be perfectly aligned with the angular momentum; when
$\gamma=0$, $\ba$ is randomly oriented with respect to $\bJ$.

As discussed at the beginning of \S \ref{sec:model}, $\bJ$
stays approximately constant between UV absorption and IR emission,
and so does $\temp{rot}$; however, the molecule internal alignment temperature
$\Tia$ may substantially change if part of the absorbed 
photon energy is transferred to rotational degrees of freedom. We
account for the uncertain energy exchange between vibrational and
rotational modes by parametrizing the grain internal alignment before
UV absorption and during IR emission respectively with $\gamma_0$ (corresponding to 
$\Tia\equiv T_0$) and $\gamma_r\leq\gamma_0$ (corresponding
to $\Tia\geq T_0$). In principle, $\gamma_r$
should be different for each IR emission feature, since it may be thought of
as the internal alignment coefficient when most of the radiation
in that band is emitted.

We now discuss another plausible choice for the parametrization of the alignment between $\ba$ and $\bJ$ prior to UV absorption. If many
collisions with hydrogen atoms occur in the interval between two UV
absorptions, and there are no other torques acting,
the grain will be driven towards ``Brownian rotation'' with
$f(J)\propto J^2$. 
If there is no vibrational-rotational energy exchange (i.e., the PAH acts
like a rigid rotator), the internal alignment temperature
$T_0$ before UV absorption will approximately equal the gas kinetic
temperature $T_{\rm gas}$. Integration of eq.~(\ref{eq:general}) with
respect to $J$ assuming $f(J)\propto J^2$ yields a probability
distribution 
\be\label{eq:align2} 
\ud P_\epsilon(\beta)=
\frac{\epsilon\,(\epsilon-1)^{1/2}\sin\beta\,\ud\beta}{2\,(\epsilon-\cos^2\beta)^{3/2}} ~~~, 
\ee
which does not depend on $T_{\rm gas}$ but only on the geometrical properties of the grain via $\epsilon\equiv I_1/(I_1-I_2)$.
 
In the following, we parametrize the disalignment between
$\ba$ and $\bJ$ by using eq.~(\ref{eq:align}) both before UV absorption
(with $\gamma_0$) and during IR emission (with
$\gamma_r\leq\gamma_0$). However, in the Appendix we present
analytic formulae for the expected degree of polarization if the
grain alignment before UV absorption can be described by
eq.~(\ref{eq:align2}), but eq.~(\ref{eq:align}) still holds during IR
emission.

 \subsection{Estimating $\gamma_0$ and $\gamma_r$}\label{sec:estimate}
In the dust model of \citet{draine_li07}, the $3-13\unit{\mu m}$
emission features are produced primarily by PAH molecules or clusters
containing between $\NC\approx 25$ and $\sim1000$ carbon
atoms, and the 17$\,\mu$m complex is mainly due to PAHs
with $\NC\approx 2000$.
For purposes of estimating rotational
kinetic energies and the  density of vibrational states, 
we will take $\NC=200$ as a representative value, with
a volume-equivalent radius $a\approx 7.5\,$\AA. We suppose that all
the emitting molecules are axisymmetric and planar, with $I_1/I_2=2$. 

Following UV absorption, the alignment between the molecule principal
axis $\ba$ and angular momentum $\bJ$, and thus the internal alignment
temperature $\Tia$, depends on the efficacy of internal energy
exchange between lattice vibrational modes and rotational modes. The
Intramolecular Vibration-Rotation Energy Transfer (IVRET) process
\citep{purcell_79}, due to imperfect elasticity of the molecule when
stressed by centrifugal and Coriolis forces, allows energy exchange
between rotation and vibrations on a timescale $\sim10^{-2}\unit{s}$
\citep{rouan_92}, much shorter than the duration $\sim1-10\unit{s}$ of
the IR emission burst. This means that, while the molecule is cooling
after UV absorption, its internal alignment temperature $\Tia$ tends
to be equal to the instantaneous vibrational (lattice) temperature
$\temp{vib}$. Following photon absorption, the lattice may be heated
up to a temperature $T_{\rm vib}\approx300 - 1500\unit{K}$,
depending on the grain size and the photon energy.  The grain
cools as IR energy is radiated, and we estimate
$\temp{vib}\approx 800\unit{K}$ when most of the 
$3.3\unit{\mu m}$ emission takes place, 
$\temp{vib}\approx 300\unit{K}$ for the $7.7\unit{\mu m}$ emission, 
$\temp{vib}\approx 200\unit{K}$ for the $11.3\unit{\mu m}$ emission,
and $\temp{vib}\approx 120\unit{K}$ for the $17\unit{\mu m}$ emission, unless the local radiation field is so intense to prevent the lattice from cooling down to such temperatures.

The internal temperature $\temp{ia}$ follows $\temp{vib}$ while the
grain is cooling, with $\Tia\approx\temp{vib}$ as long as the
vibrational energy levels are sufficiently closely spaced to allow 
energy transfer between vibrations and rotation.\footnote{If
$\hat{\bf a}_1$ and ${\bf J}$ are not parallel and $I_2\neq I_3$, the
molecule nutates. Because of nutation, the rotational motion around
$\ba$ is only quasiperiodic, and the centrifugal and Coriolis stresses
responsible for the IVRET process will have Fourier components over a
range of frequencies around the mean rotation rate $\omega_{\rm rot}$
(i.e., the rotation rate averaged over the nutation period). This
facilitates the coupling of rotation and vibrations via the IVRET
process.}  When the separation $\Delta E$ of vibrational levels
exceeds $\sim\hbar\,\omega_{\rm rot}$,
the IVRET process ceases to operate and the
rotational modes decouple from the lattice.  The density of states can
be calculated using a model normal mode spectrum and the
Beyer-Swinehart algorithm \citep{draine_li01}: for
$N_{{\msc{C}}}=200$, the density of states at vibrational energy $E/hc
\approx 250\unit{cm^{-1}}$ is $hc\, \ud N/\ud E\approx 1/ (0.18
\unit{cm^{-1}})$.  This should still allow energy exchange in quanta
$\Delta E/hc \sim (\omega_{\rm rot}/2\pi c) \approx 0.33
\unit{cm}^{-1}$ for a grain spinning at $\omega_{\rm rot}/2\pi
\approx10\unit{GHz}$, so that the rotational and vibrational modes
will decouple only when the lattice vibrational energy has dropped
below $E/hc\approx 250\,{\rm cm}^{-1}$.  For our model PAH, this is
the average energy if it were in contact with a
$\sim65\unit{K}$ heat bath, which suggests $T_{0}\approx
65\unit{K}$ to describe alignment of $\hat{\bf a}_1$ with ${\bf J}$
prior to photon absorption. As we discuss below, this is actually a
lower limit for $T_0$, which is presumably attained in cold
interstellar clouds, whereas in bright photodissociation regions like
the Orion Bar the radiation field is so dense that PAH grains stay
always hotter.

As anticipated in \S \ref{sec:align}, we assume that the grain
disalignment between principal axis $\ba$ and angular momentum $\bJ$
may be described by the probability distribution in \eq{align} with
``internal alignment'' coefficient $\gamma_0$ 
just before UV absorption, and $\gamma_r$ during IR emission. The ratio
$T_{\rm rot}/\Tia$ 
depends on environmental conditions, hence so does
$\gamma$ in \eq{gam_def}. We now discuss four possible cases,
summarized in Table \ref{tab:cases}; for each case, we compute the
value of $\gamma_0$ prior to absorption of starlight photons, and the
values of $\gamma_r$ during emission of radiation in the 3.3, 7.7,
11.3 and 17$\,\mu$m bands.

 \begin{deluxetable}{l c c c c}
\tablecaption{\label{tab:cases}Selected Cases for a Disk Molecule with $\NC=200$}
\tablehead{
  \colhead{Parameter} &
  \colhead{case (a)} &
  \colhead{case (b)} &
  \colhead{case (c)} &
  \colhead{case (d)} \\
  \colhead{}       &
  \colhead{CNM}    &
  \colhead{Orion Bar} &
  \colhead{STR} &
  \colhead{$\hat{\bf a}_1\parallel{\bf J}$} }
\startdata
$T_{\rm rot}\,$(K)         & 80  & 220 & 1000 & $\infty$ \\
$T_0\,$(K)                 & 65  & 150 & 60   & -- \\
$\gamma_0$                 & 1.2 & 1.5 & 17  & $\infty$ \\
$\gamma_r(3.3\mu{\rm m})$  & 0.10 & 0.28 & 1.3  & $\infty$ \\
$\gamma_r(7.7\mu{\rm m})$  & 0.27 & 0.73 & 3.3  & $\infty$ \\
$\gamma_r(11.3\mu{\rm m})$ & 0.40 & 1.1 & 5.0  & $\infty$ \\
$\gamma_r(17\mu{\rm m})$   & 0.67 & 1.5 & 8.3    & $\infty$ \\
\hline
$p_\star^{\parallel}(\pi/2)$, $3.3\mu{\rm m}$ (\%)  
                 &  $0.02$ & $0.06$ & $1.29$ & $7.69$ \\
$p_\star^{\parallel}(\pi/2)$, $7.7\mu{\rm m}$ (\%) 
                 &  $0.05$ & $0.17$ & $3.29$ & $7.69$ \\
$p_\star^{\perp}(\pi/2)$, $11.3\mu{\rm m}$ (\%)
                 & $-0.14$ & $-0.53$ & $-8.56$ & $-14.29$ \\
$p_\star^{\perp}(\pi/2)$, $17\mu{\rm m}$ (\%)
                 & $-0.25$ & $-0.73$ & $-10.54$ & $-14.29$
\enddata
\end{deluxetable}

\subsubsection{Case {\rm(a)}: CNM}\label{sec:cas_a}
In H~I clouds with gas kinetic temperature $T_{\rm gas}\approx 100\,$K
(the ``Cold Neutral Medium'', or CNM), the rotational excitation model
by \citet{draine_lazarian98} predicts that a PAH with $\NC=200$ should
be spinning with rotational temperature $\temp{rot}\approx80$K, which
corresponds to a rotation frequency $\omega_{\rm rot}/2\pi\approx
10\unit{GHz}$. Since the molecule's angular momentum stays
approximately constant between UV absorption and IR emission, we may
assume $T_{\rm rot}\approx80\unit{K}$ during the whole process. If
$T_0\approx65\unit{K}$ prior to UV absorption, as discussed above, the
internal alignment coefficient of our model PAH will be
$\gamma_0\approx1.2$; the values of $\gamma_r$ for different IR
emission bands are listed in Table \ref{tab:cases}.

We note that, if gas collisions were frequent enough to drive the
emitting grains towards ``Brownian rotation'' between two subsequent
UV absorptions, then the probability distribution in \eq{align2}
should be used instead of \eq{align} to describe the internal
alignment prior to UV absorption, whereas \eq{align} may still be used
during IR emission with an appropriate choice for $\gamma_r$. However,
as discussed in the Appendix, identical polarization results may be
obtained in the current formalism if we employ an ``effective''
internal alignment coefficient $\gamma_0\approx1.0$ before UV
absorption, similar to the value $\gamma_0=1.2$ adopted for case
(a).

\subsubsection{Case {\rm(b)}: The Orion Bar PDR} \label{sec:cas_b}
The Orion Bar, illuminated by the Trapezium stars, is an example of a
bright photodissociation region (PDR), with strong PAH emission at
3.3$\;\micron$ \citep{Tielens+Meixner+vanderWerf+etal_1993}.  Physical
conditions in the photodissociation zone have been discussed by
\citet{allers_05}, who estimate
$n_{\msc{H}}\approx7\times10^4\unit{cm^{-3}}$,
$\temp{gas}\approx1000\unit{K}$, and $\chi\approx 3\times10^4$, where
$n_{\msc{H}}$ is the hydrogen number density and $\chi$ is the specific
radiation energy density at 1000\,\AA\ relative to the value in the
local interstellar radiation field.  For these conditions, the
rotational excitation model of \citet{draine_lazarian98} yields a
rotational temperature $\temp{rot}\approx220\unit{K}$ for a PAH with
$\NC=200$.

In the intense radiation field of a bright PDR, the vibrational energy
of our model PAH is always large enough (and the vibrational levels
sufficiently closely spaced) that the internal alignment temperature
$\temp{ia}$ roughly equals the instantaneous vibrational temperature
$\temp{vib}$ at any time. For a PAH molecule with $\NC=200$ in a
radiation field with $\chi\approx3\times10^4$, the lattice can hardly
cool below $\temp{vib}\approx150\unit{K}$ \citep{draine_li01}, so that
in a PDR like the Orion Bar $T_0\approx150\unit{K}$ may be appropriate
prior to UV absorption. The corresponding internal alignment
coefficient will be $\gamma_0\approx1.5$ for
$\temp{rot}\approx220\unit{K}$.  Values of $\gamma_r$ for different
emission bands are given in Table \ref{tab:cases}; since $\temp{vib}$
never drops below $\approx150\unit{K}$, we take
$\temp{ia}\approx150\unit{K}$ as the characteristic internal alignment
temperature for emission in the $17\unit{\mu m}$ band.

\subsubsection{Case {\rm(c)}: Suprathermal Rotation (STR)}\label{sec:cas_c}
As discussed in \S \ref{sec:cas_a} and \ref{sec:cas_b}, PAH molecules
in the CNM should have moderately sub-thermal rotation rates
($\temp{rot}\lesssim\temp{gas}$), whereas $\temp{rot}\ll\temp{gas}$ in
bright PDRs like the Orion Bar. Observations of microwave emission
from spinning grains in the diffuse ISM
\citep[e.g.,][]{Dobler+Finkbeiner_2008a,
Dobler+Draine+Finkbeiner_2009} and in PDRs
\citep[e.g.,][]{Casassus+Dickinson+Cleary+etal_2008} appear to be
consistent with these estimates.

However, it is of interest to explore the possibility that some PAHs
might be subject to systematic torques which could spin them up to
higher rotation rates. For example, inelastic collisions between a
grain and hydrogen atoms may be followed by ejection of H$_2$
molecules when a C-H bond is broken through photo- or
thermo-dissociation triggered by the absorption of UV photons. If
recombination of H atoms preferentially occurs at a few catalytic
centers spread over the grain surface, and the H$_2$ ejection is
systematically asymmetric, the resulting torque may significantly spin
up the PAH molecule.

For case (c) we assume $T_{\rm
rot}\sim10\,\temp{gas}\approx1000\unit{K}$ for PAHs in H~I clouds,
corresponding to a rotation frequency $\omega_{\rm rot}/2
\pi\approx30\unit{GHz}$ for our model grain with $\NC=200$ carbon
atoms.  For PAHs spinning at $30\unit{GHz}$, efficient energy transfer
between vibrations and rotation via the IVRET process is suppressed,
due to insufficient density of vibrational states, when the lattice
cools below $\temp{vib}\approx60\unit{K}$, hence we take
$T_0\approx60\unit{K}$ as the internal alignment temperature prior to
UV absorption, giving an internal alignment coefficient
$\gamma_0\approx17$.  Values of $\gamma_r$ for different PAH emission
bands are listed in Table \ref{tab:cases}.

\subsubsection{Case {\rm(d)}: $\ba\;||\;\bJ$}\label{sec:cas_d}
The physics of internal relaxation in cold, spinning grains is not
well understood, and perhaps there is a slow process that couples the
rotational degrees of freedom to the lattice down to very low
temperatures, regardless of the lower limit imposed by the density of
vibrational states.  Since in H~I clouds the interval between two
subsequent photon absorptions is long ($\sim 10^7\unit{s}$), there is
a lot of time for a weak process to act.  Such a process would lead to
$\gamma_0 \gg 1$, and we discuss $\gamma_0=\infty$ as a limiting case.

Also, if the coupling between rotation and vibrations after UV
absorption took place on timescales much longer than the duration of
the IR emission burst ($\sim1-10\unit{s}$), the internal alignment
temperature during IR emission would stay roughly equal to the value
before UV absorption. If $\gamma_0\gg1$, this would imply
$\gamma_r\gg1$ as well. We therefore consider $\gamma_0=\infty$ and
$\gamma_r=\infty$ as the limiting case in which $\ba$ is perfectly
aligned with $\bJ$ both before UV absorption and during IR emission.

\section{Polarization from PAH{\tiny s} illuminated by a star or a disk galaxy}\label{sec:results}
We now compute the degree of polarization expected for in-plane and
out-of-plane vibrational modes of PAH molecules when anisotropically
illuminated by a star or a disk galaxy. We assume, for the sake of
simplicity, that the angular momentum of emitting grains is randomly
oriented in space, although it may be partially aligned by the
interstellar magnetic field. Also, we suppose that the illuminating
disk galaxy has a uniform surface-brightness profile. However, the
formalism developed in \S \ref{sec:model} can be applied without such
restrictions, and we discuss below how to relax these two
assumptions. The grain alignment between $\ba$ and $\bJ$ is described
via the probability distribution in \eq{align} both before UV
absorption ($\ud P_{\gamma_0}$) and during IR emission ($\ud
P_{\gamma_r}$), with ``internal alignment'' coefficients $\gamma_0$
and $\gamma_r$ respectively.

For a population of PAH
molecules illuminated by a point source, 
the emission polarized along a direction $\hb w$ (= either
$\hat{\b{u}}$ or $\hat{\b{v}}$ -- see \fig{geom}\pan{a}) for in-plane
($\parallel$) and out-of-plane ($\perp$) modes is
\begin{eqnarray}\label{eq:polstar}
I_{\star,
w}^{\parallel,\perp}(\alpha,\gamma_0,\gamma_r)&\propto\!\!&\int_{-1}^{1}\!\!\!\ud\cos\theta\int_0^{2\pi}\!\!\!\ud\varphi\int_0^\pi\!\!\!\ud
P_{\gamma_0}(\beta_0)\int_0^\pi\!\!\!\ud
P_{\gamma_r}(\beta_r)\nonumber\\ &&\cross\,\bar
A_\star(\theta,\beta_0)\,\bar
F_w^{\parallel,\perp}(\theta,\varphi,\beta_r,\alpha)~,
 \end{eqnarray}
where $\bar A_\star$ and $\bar F_w^{\parallel,\perp}$ are the
absorption and emission cross sections computed in \S \ref{sec:abs} and \S \ref{sec:em}
respectively. If the grain angular momentum is not isotropically
oriented in space, it is easy to incorporate the corresponding
$(\theta,\varphi)$-distribution function into the above integral. The
polarized emission from PAHs above the center
of a uniform-brightness disk galaxy is
\begin{eqnarray}\label{eq:polgal}
I_{\mathrm{gal}, w}^{\parallel,\perp}(\alpha,\gamma_0,\gamma_r,\omega)&\propto&\!\!\int_{-1}^{1}\!\!\!\!\ud\cos\theta\!\int_0^{2\pi}\!\!\!\!\!\!\ud\varphi\!\int_0^\pi\!\!\!\ud P_{\gamma_0}(\beta_0)\!\int_0^\pi\!\!\!\ud P_{\gamma_r}(\beta_r)\nonumber\\&&\!\!\!\!\!\!\!\!\!\!\!\!\!\!\!\!\!\!\!\!\!\!\!\!\!\!\!\!\!\!\cross\int_{\cos\omega}^{1}\!\!\!\!\!\!\!\frac{\ud\cos\theta'}{\Omega} \bar A_{\rm{gal}}(\theta,\beta_0,\theta')\,\bar F_w^{\parallel,\perp}(\theta,\varphi,\beta_r,\alpha)~, 
 \end{eqnarray}
where $\Omega=2\pi(1-\cos\omega)$ is the solid angle subtended by the
galactic disk as seen from the emitting molecules. A generic
axisymmetric surface-brightness profile $B(\theta')$ can be easily
included in the previous expression.

The resulting degree of polarization for in-plane and out-of-plane modes is respectively
 \begin{equation}\label{eq:pol_def}
 p^{\parallel,\perp}\equiv\frac{I_u^{\parallel,\perp}-I_v^{\parallel,\perp}}{I_u^{\parallel,\perp}+I_v^{\parallel,\perp}}~~~,
 \end{equation}
and will therefore be positive for emission polarized along $\hat{\mathbf{u}}$ and negative for polarization along $\hat{\mathbf{v}}$.

\subsection{Polarization from PAHs illuminated by a star}\label{sec:starpol}
By integrating eq.~(\ref{eq:polstar}), we obtain
the degree of
polarization for in-plane and out-of-plane modes from a population of
PAH molecules illuminated by a star (or a generic point source) 
\begin{eqnarray}\label{eq:star1}
p^\parallel_\star(\alpha,\gamma_0,\gamma_r)&=&\frac{3\,
\sin^2\alpha}{640\,h(\gamma_0)h(\gamma_r)+3\, \cos^2\alpha-1}~,\\
p^\perp_\star(\alpha,\gamma_0,\gamma_r)&=&\frac{-3\,
\sin^2\alpha}{320\,h(\gamma_0)h(\gamma_r)-3\,
\cos^2\alpha+1}~,\label{eq:star2}
\end{eqnarray}
where $\alpha$ is the angle between the line of sight and the
illumination direction (see \fig{geom}\pan{a}), and
\begin{eqnarray}\label{eq:h_def}
g_1(\gamma)&\equiv&\sqrt{\pi}\, \gamma\,
\mathrm{erfi(\sqrt{\gamma})}~~~,\nonumber\\
g_2(\gamma)&\equiv&6\,\sqrt{\gamma}\,e^\gamma-(3+2\gamma)\sqrt{\pi}\,\mathrm{erfi(\sqrt{\gamma})}~~~,\nonumber\\
h(\gamma)&\equiv&g_1(\gamma)/g_2(\gamma)~~~.
\end{eqnarray}
 The function $h(\gamma)$
monotonically decreases from $h(0)\rightarrow\infty$
to $h(\gamma\rightarrow\infty)=1/4$.  Since
$h(0)\rightarrow\infty$, if the grain principal axis
of inertia is randomly oriented with respect to the angular momentum
either before or after UV absorption, the emitted radiation is
completely unpolarized. Since $h(\gamma)>1/4$, we have
$p^{\parallel}_\star\geq0$ and $p^{\perp}_\star\leq0$ for any choice
of $\gamma_0,\gamma_r$ and $\alpha$. According to the definition in
eq.~(\ref{eq:pol_def}), this means that 
in-plane and out-of-plane modes are polarized respectively along $\hat{\b u}$ and
along $\hat{\b v}$, confirming Leger's (1988)
prediction that the polarization direction for in-plane and
out-of-plane modes is respectively orthogonal and parallel to the
plane-of-sky projection of the illumination direction (see \fig{geom}\pan{a}).

\begin{figure}[htbp]
\centering 
\includegraphics[width=0.5\textwidth]{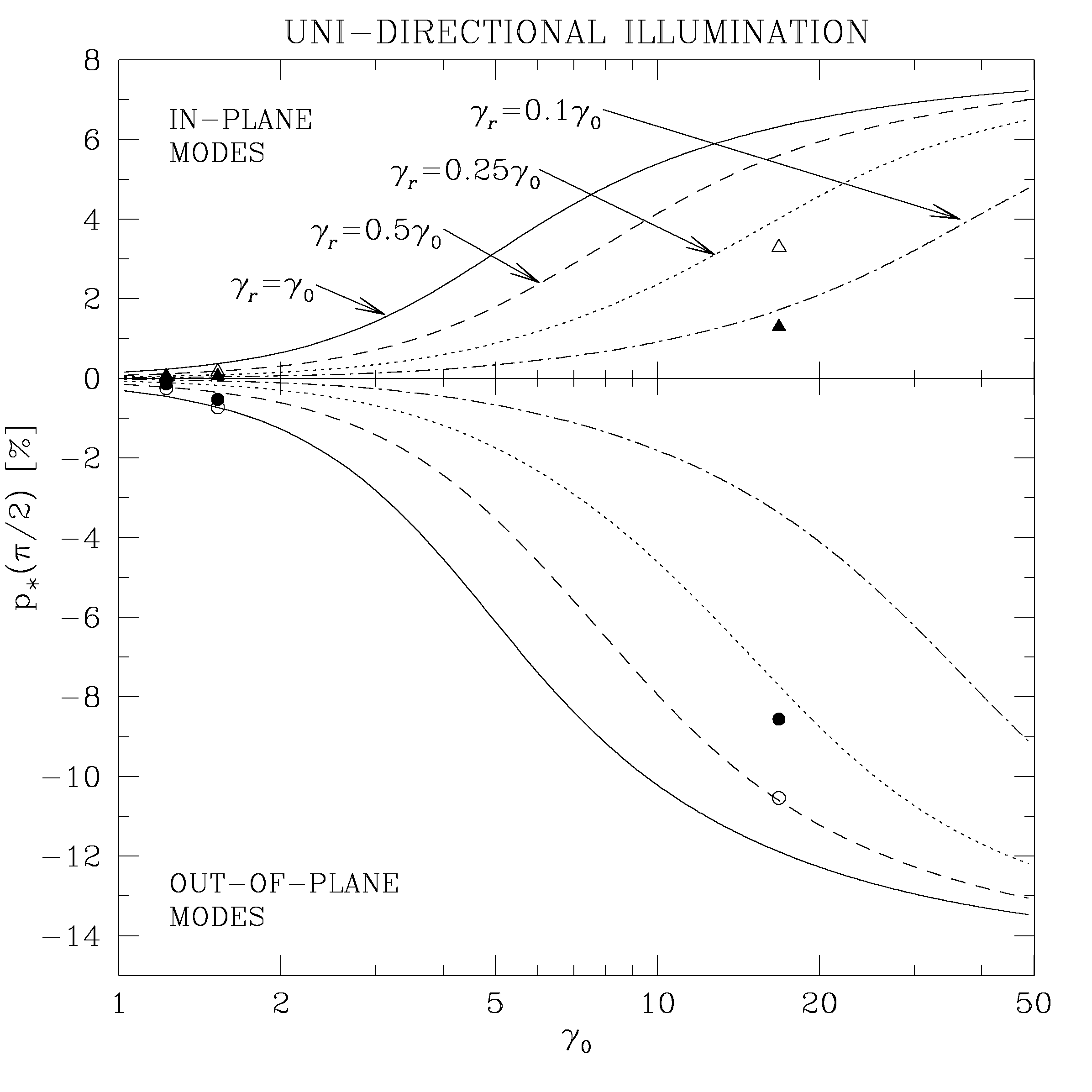}
\caption{For PAHs illuminated by a star, dependence of the
polarization $p_\star$ on $\gamma_0$, for optimal viewing geometry
($\alpha=\pi/2$). The different lines correspond to different values
of the ratio $\gamma_r/\gamma_0$: $\gamma_r=\gamma_0$ (solid),
$\gamma_r=0.5\,\gamma_0$ (dashed), $\gamma_r=0.25\,\gamma_0$ (dotted),
and $\gamma_r=0.1\,\gamma_0$ (dot-dashed). The polarization values
reported in Table \ref{tab:cases} are shown as solid triangles
(3.3$\unit{\mu m}$), open triangles (7.7$\unit{\mu m}$), solid circles
(11.3$\unit{\mu m}$), and open circles (17$\unit{\mu m}$, assumed to
arise from out-of-plane modes), with different values of $\gamma_0$
corresponding to case (a) ($\gamma_0=1.2$), case (b) ($\gamma_0=1.5$)
or case (c) ($\gamma_0=17$).\\}
\label{fig:star} 
\end{figure}

\begin{figure}[htbp]
\centering 
\includegraphics[width=0.5\textwidth]{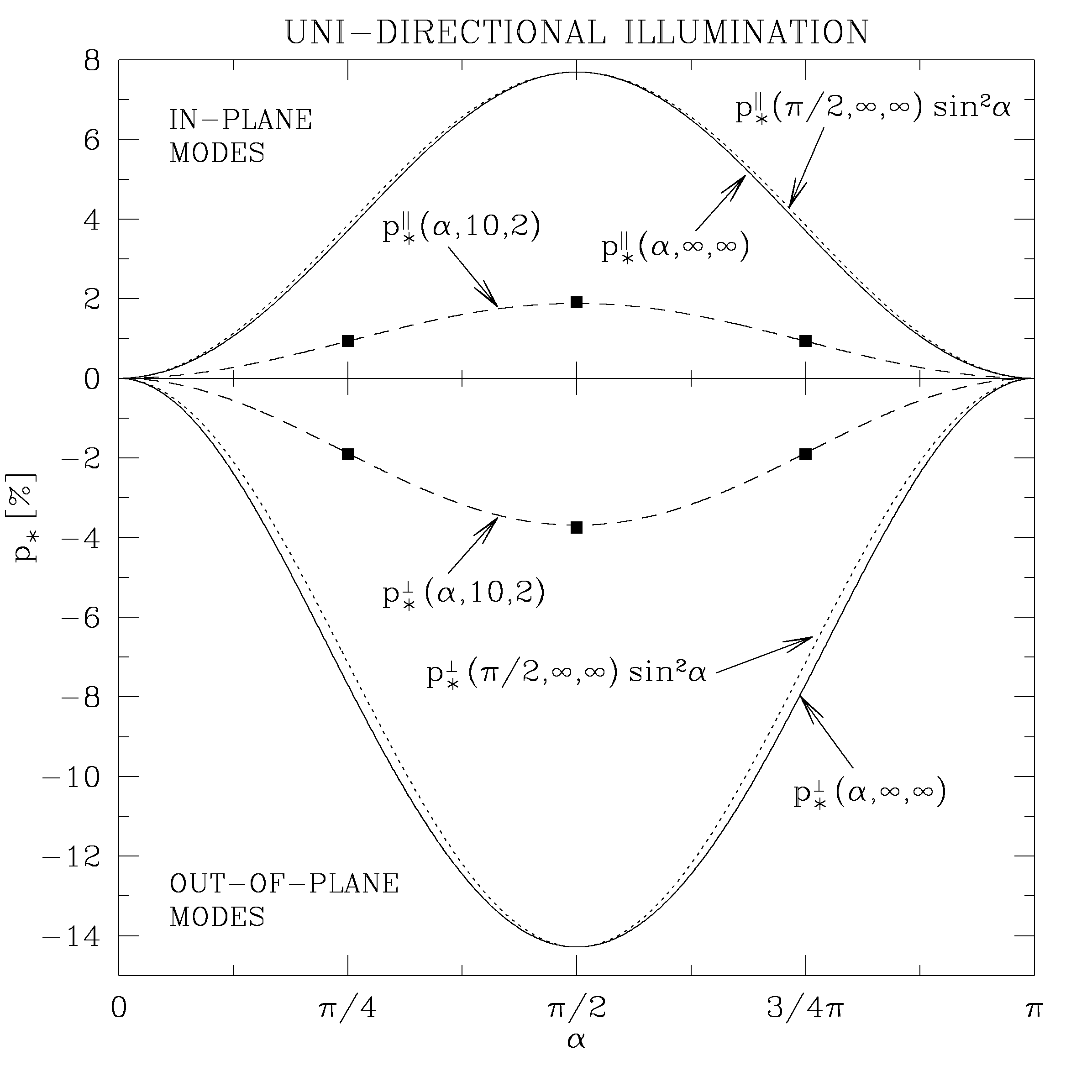}
\caption{For PAHs illuminated by a star, dependence of the 
polarization $p_\star(\alpha,\gamma_0,\gamma_r)$ 
on the viewing angle $\alpha$.
Dashed line: polarization for $(\gamma_0,\gamma_r)=(10,2)$, similar
to case (c) in \S\ref{sec:estimate}.  The polarization values quoted
by \citet{leger_88} for a disk molecule are shown as solid squares,
and are seen to match the values calculated for
$(\gamma_0,\gamma_r)=(10,2)$.
Solid line: polarization for perfect internal alignment
$(\gamma_0,\gamma_r)=(\infty,\infty)$
(see eqs.~(\ref{eq:star3})-(\ref{eq:star4})), corresponding to case (d) in
\S\ref{sec:estimate}.
The dotted line is
$p^{\parallel,\perp}_\star(\pi/2)\,\sin^2\alpha$ for
$(\gamma_0,\gamma_r)=(\infty,\infty)$, which closely
approximates the dependence on $\alpha$ of
eqs.~(\ref{eq:star3})-(\ref{eq:star4}).\\}
\label{fig:approx} 
\end{figure}

For fixed $\gamma_0$ and $\gamma_r$, eqs.~(\ref{eq:star1}) and
(\ref{eq:star2}) show that the IR emission bands should be maximally
polarized when $\alpha=\pi/2$, i.e., the line of sight to the emitting
PAHs is orthogonal to the star-molecule direction. For this optimal
viewing geometry, Table \ref{tab:cases} reports the degree of
polarization expected for the 3.3, 7.7, 11.3 and 17$\unit{\mu}$m
emission features in different environmental conditions, as discussed
in \S\ref{sec:estimate}; we have assumed the $17\unit{\mu m}$ band to
arise from out-of-plane modes.  Figure \ref{fig:star} shows the
dependence of $p^\parallel_\star(\pi/2)$ and $p^\perp_\star(\pi/2)$ on
the internal alignment coefficient $\gamma_0$, with the different
curves corresponding to different values of the ratio
$\gamma_r/\gamma_0$.
Because
$\gamma_r\leq\gamma_0$, the region to the left of the solid lines is
excluded. In most cases, the dependence on the viewing angle $\alpha$
in the denominator of eqs.~(\ref{eq:star1})-(\ref{eq:star2}) can be
neglected, hence $p^{\parallel}_\star(\alpha,\gamma_0,\gamma_r)\approx
p^{\parallel}_\star(\pi/2,\gamma_0,\gamma_r)\,\sin^2\alpha$, and
similarly for $p^\perp_\star$ (compare solid and dotted lines in
\fig{approx}).

For the extreme case of perfect alignment between $\ba$ and $\bJ$
during both UV absorption ($\gamma_0=\infty$) and IR emission
($\gamma_r=\infty$), corresponding to case (d) in
\S\ref{sec:estimate}, eqs.~(\ref{eq:star1}) and (\ref{eq:star2}) may
be simplified:
\begin{eqnarray}
p^\parallel_\star(\alpha)=\frac{\sin^2\alpha}{13+\cos^2\alpha}~,\label{eq:star3}\\
p^\perp_\star(\alpha)=\frac{-\sin^2\alpha}{7-\cos^2\alpha}~.\label{eq:star4}
\end{eqnarray}

\subsection{Polarization from PAHs illuminated by a uniform-brightness disk galaxy}
\begin{figure}[htbp]
\centering 
\includegraphics[width=0.5\textwidth]{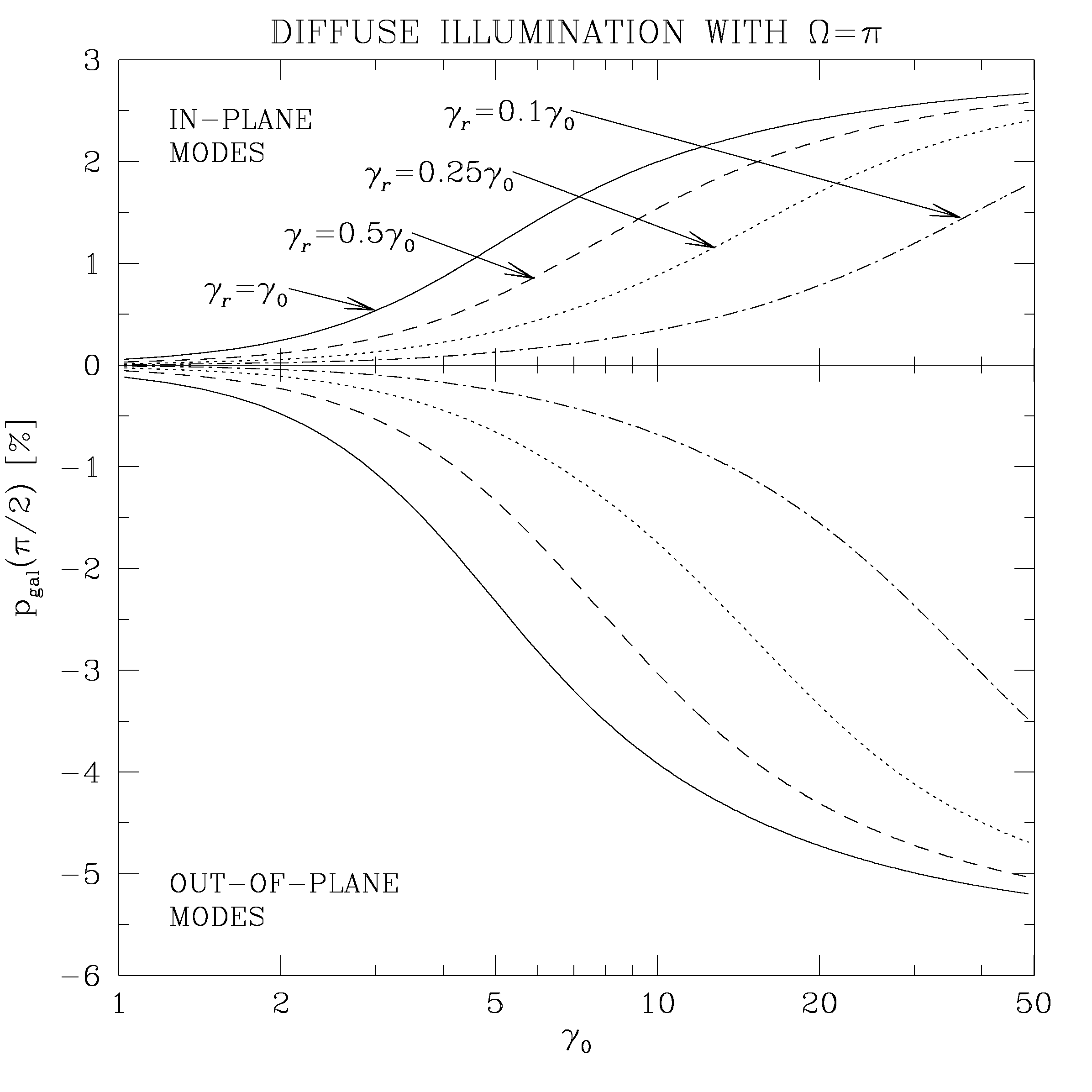}
\caption{For PAHs illuminated by a disk galaxy with $\Omega=\pi$,
dependence of the polarization $p_{\rm gal}$ on $\gamma_0$, for optimal
viewing geometry ($\alpha=\pi/2$). The different lines correspond to
different values of the ratio $\gamma_r/\gamma_0$, as explained in the
caption of \fig{star}.\\}
\label{fig:gal} 
\end{figure}

The degree of polarization for in-plane and out-of-plane modes from a
population of PAH molecules above the center of a uniform-brightness
disk galaxy is obtained from \eq{polgal}:
\begin{eqnarray}\label{eq:galaxy1}
\!\!\!\!\!\!\!p^\parallel_{\mathrm{gal}}(\alpha,\gamma_0,\gamma_r,\omega)&=&\frac{3\,
\sin^2\alpha}{1280\frac{h(\gamma_0)h(\gamma_r)}{(\cos^2\omega+\cos\omega)}+3\,
\cos^2\alpha-1}~,~\,\\
\!\!\!\!\!\!\!p^\perp_\mathrm{gal}(\alpha,\gamma_0,\gamma_r,\omega)&=&\frac{-3\,
\sin^2\alpha}{640\frac{h(\gamma_0)h(\gamma_r)}{(\cos^2\omega+\cos\omega)}-3\,
\cos^2\alpha+1}~,
\label{eq:galaxy2}
\end{eqnarray}
where $\Omega=2\pi(1-\cos\omega)$ is the angle subtended by the galaxy
as seen from the emitting grains. The limit $\Omega\rightarrow0$
recovers the case of a point-like illuminating source in
eqs.~(\ref{eq:star1}) and (\ref{eq:star2}), whereas for an infinite
disk ($\Omega\rightarrow2\pi$) the degree of polarization tends to
zero.  Figure \ref{fig:gal} shows, for the most favorable viewing
geometry ($\alpha=\pi/2$), the dependence of the degree of
polarization on $\gamma_0$ for PAHs illuminated by a disk galaxy with
$\Omega=\pi$; the different curves correspond to different values of
$\gamma_0/\gamma_r$, as explained in \S\ref{sec:starpol}.  For diffuse
illumination with $\Omega=\pi$, we find levels of polarization that
are $\sim40\%$ of the values found for uni-directional illumination.

The extreme case of grains with perfect internal alignment
($\ba\parallel\bJ$ during both UV absorption and IR emission, case (d) in \S\ref{sec:estimate}) yields
\begin{eqnarray}\label{eq:galaxy1bis}
\!\!\!p^\parallel_{\mathrm{gal}}(\alpha,\omega)&=&\frac{3\,
\sin^2\alpha}{80(\cos^2\omega+\cos\omega)^{-1}+3\,
\cos^2\alpha-1}~,\,\\
\!\!\!p^\perp_\mathrm{gal}(\alpha,\omega)&=&\frac{-3\,
\sin^2\alpha}{40(\cos^2\omega+\cos\omega)^{-1}-3\,
\cos^2\alpha+1}~.\label{eq:galaxy2bis}
\end{eqnarray}

\section{Discussion and summary}\label{sec:concl}
We have obtained analytic formulae for
the degree of polarization expected for the 3.3, 6.2,
7.7, 8.6, 11.3, 12.7, 16.4, and 17$\,\um$ 
emission features when the emitting
PAHs are anisotropically illuminated by a source of UV
photons. We model PAH grains as planar molecules with in-plane and
out-of-plane vibrational dipoles. Vibrational modes oscillating in the molecular plane
(responsible for the 3.3, 6.2, 7.7, and 8.6$\unit{\mu m}$ emission bands) are
polarized perpendicular to the plane-of-sky projection of the
illumination direction, whereas for out-of-plane modes (responsible
for the 11.3 and 12.7$\unit{\mu m}$ features) the polarization is along the
source-molecule direction, as originally pointed out by \citet{leger_88}.
The fact that the predicted polarization directions for in-plane and out-of-plane modes are orthogonal  provides a check on the contribution
to the polarization from linear dichroism by aligned foreground dust,
as this would be expected to produce
polarization in a single plane.
It also means that polarization measurements could be used to diagnose the
character (in-plane or out-of-plane) of the modes responsible for the poorly-characterized 16.4 and 17$\,\mu$m emission
features.

Our analytic formulae show explicitly how the degree of polarization
depends on the viewing geometry (i.e., the angle $\alpha$ between the
line of sight and the illumination direction) and the ``internal
alignment'' coefficients $\gamma_0$, describing the alignment between
the grain principal axis $\ba$ and angular momentum $\bJ$ just before
UV absorption, and $\gamma_r$, characterizing the internal alignment
during the brief period of IR emission following UV excitation. The
degree of polarization is maximal for $\alpha=\pi/2$ and, at fixed
$\alpha$, it increases with $\gamma_0$ and $\gamma_r$, as the
principal axis and angular momentum become more closely aligned.  We
have discussed both the case of a point-like illuminating source,
which may be applied to reflection nebulae and PDRs such as the Orion
Bar, and of an extended source (e.g., dust above a disk galaxy like
NGC~891 or M82). The degree of polarization for uni-directional
illumination is higher than for diffuse illumination, all else being
equal; however, if the galaxy's surface-brightness profile is strongly
peaked at the galactic center, the galaxy would resemble a point
source, with polarization levels comparable to the case of
uni-directional illumination.

The value of $\gamma_r$ is expected to increase with increasing
wavelength of the emission feature, since such a band will be mostly
emitted when the grain is cooler and $\ba$ and $\bJ$ are expected to be more
closely aligned. Therefore, when comparing two IR emission features
arising from modes of the same character (either in-plane or
out-of-plane), the band with longer wavelength should be more strongly
polarized.  Table \ref{tab:cases} reports, in case of uni-directional
illumination and optimal viewing geometry ($\alpha=\pi/2$), the
predicted degree of polarization for the 3.3, 7.7, 11.3, and
17$\,\mu$m emission features, for four exemplary choices of
environmental conditions.\footnote{
    The polarization
    results in Table \ref{tab:cases} assume that the PAH angular momenta
    are randomly oriented in space; if the grains are
    partially oriented, e.g., by the interstellar magnetic field, we would
    expect a higher degree of polarization.}

For dust in the CNM, significant disalignment between $\ba$ and $\bJ$
($\gamma_0\approx 1$, $\gamma_r\lesssim0.5\,\gamma_0$) is expected to strongly
suppress the intrinsic polarization of the IR bands, with the
3.3$\,\mu$m emission feature predicted to be polarized by only
$\approx0.02\%$ (for our estimated $\gamma_0=1.2$ and
$\gamma_r=0.10$), presumably too small to distinguish from
polarization due to foreground linear dichroism.  Longer wavelength
features are expected to be more strongly polarized (e.g., $-0.14\%$
for the 11.3$\,\mu$m feature, with $\gamma_r=0.40$), but the
polarization levels are still observationally challenging.
These conditions may
apply to PAHs above the disk of an edge-on galaxy, such as NGC~891 or
M82; however, the polarization degrees quoted above are computed for
the case of uni-directional illumination, which is realized only if
the galaxy's luminosity is strongly concentrated at its center.
If the galaxy appears to the emitting grains as an
extended source, the PAH polarization will be even smaller.

The Orion Bar PDR is expected to be somewhat more favorable for the
PAH internal alignment because of higher molecular rotation rates, but
even there the polarization of the 3.3$\,\mu$m feature is predicted to
be only $\approx0.06\%$ for our estimated $(\gamma_0,\gamma_r)=(1.5,
0.28)$.  If the polarization $0.86\pm0.28\,\%$ measured by
\citet{sellgren_88} for the 3.3$\,\mu$m emission feature is correct,
it implies that we underestimated the internal alignment of emitting
PAHs both before UV absorption ($\gamma_0$) {\it and} during IR
emission ($\gamma_r$).  A polarization of $0.86\%$ for optimal viewing
angle $\alpha=\pi/2$ would require, e.g.,
$(\gamma_0,\gamma_r)=(2.3,2.3)$, $(3.0, 1.8)$, $(\infty, 0.81)$, etc.:
all solutions must have $\gamma_r\leq\gamma_0$, hence $\gamma_0 \geq
2.3$ {\it and} $\gamma_r \geq 0.81$.  Such good internal alignment is
not expected unless the smallest PAHs are rotating suprathermally.
The polarization at $11.3\,\mu$m is predicted to be considerably
larger than the polarization at $3.3\,\mu$m (see Table
\ref{tab:cases}).  Additional polarization measurements of the Orion
Bar at $3.3$ and 11.3$\,\mu$m are needed.  If the large intrinsic
polarization detected by \citet{sellgren_88} at $3.3\,\micron$ is
confirmed, it will call into question current thinking with regard to
the rotational dynamics of PAHs in PDRs, since it seems to require at
least moderately suprathermal rotation rates.

Comparison of our analytic results with polarization measurements may
provide useful constraints on the geometrical and dynamical properties
of PAH molecules, concerning in particular: \textit{i}) their
planarity, which has been assumed \textit{a priori} in this work,
although it is still uncertain, especially for larger PAHs;
\textit{ii}) the efficacy of internal relaxation processes that can
exchange energy between vibrational and rotational modes;
\textit{iii}) the possibility of systematic torques that may spin the
grains up to suprathermal rotation; and \textit{iv}) a determination,
via the polarization direction, of the character of the vibrational
modes contributing to a given IR feature, which will be of particular
interest for the poorly-characterized 16.4 and 17$\,\mu$m bands.

If the polarization is too small to be detected, the interpretation
will not be clear: it could be the result of PAH non-planarity, or a
mixture of in-plane and out-of-plane contributions to the mode (this
may be an issue for the 16.4 and 17$\,\mu$m features), or (more
likely) it could simply derive from poor alignment of the grain
principal axis $\hat{\bf a}_1$ and angular momentum $\bJ$.  However,
if a high degree of polarization is observed, it will imply both that
PAHs are planar {\it and} that their principal axis is well aligned
with the angular momentum.  This would be valuable information
concerning the nature of PAHs and their rotational dynamics.

\acknowledgements BTD thanks Chris Packham and Charles Telesco for
helpful discussions on the possibility of polarization of the PAH
features. We are grateful to R.H. Lupton for the availability of the
SM graphics program.
This research was supported in part by NSF grant \mbox{AST-0406883}.

\clearpage
\bibliography{pol}

\begin{thebibliography}{16}
\expandafter\ifx\csname natexlab\endcsname\relax\def\natexlab#1{#1}\fi

\bibitem[{{Allamandola} {et~al.}(1985){Allamandola}, {Tielens}, \&
  {Barker}}]{allamandola_85}
{Allamandola}, L.~J., {Tielens}, A.~G.~G.~M., \& {Barker}, J.~R. 1985, \apjl,
  290, L25

\bibitem[{{Allers} {et~al.}(2005){Allers}, {Jaffe}, {Lacy}, {Draine}, \&
  {Richter}}]{allers_05}
{Allers}, K.~N., {Jaffe}, D.~T., {Lacy}, J.~H., {Draine}, B.~T., \& {Richter},
  M.~J. 2005, \apj, 630, 368

\bibitem[{{Casassus} {et~al.}(2008){Casassus}, {Dickinson}, {Cleary},
  {Paladini}, {Etxaluze}, {Lim}, {White}, {Burton}, {Indermuehle}, {Stahl}, \&
  {Roche}}]{Casassus+Dickinson+Cleary+etal_2008}
{Casassus}, S., {Dickinson}, C., {Cleary}, K., {Paladini}, R., {Etxaluze}, M.,
  {Lim}, T., {White}, G.~J., {Burton}, M., {Indermuehle}, B., {Stahl}, O., \&
  {Roche}, P. 2008, \mnras, 391, 1075

\bibitem[{{Dobler} {et~al.}(2008){Dobler}, {Draine}, \&
  {Finkbeiner}}]{Dobler+Draine+Finkbeiner_2009}
{Dobler}, G., {Draine}, B.~T., \& {Finkbeiner}, D.~P. 2008, astro-ph/0811.1040

\bibitem[{{Dobler} \& {Finkbeiner}(2008)}]{Dobler+Finkbeiner_2008a}
{Dobler}, G. \& {Finkbeiner}, D.~P. 2008, \apj, 680, 1222

\bibitem[{{Draine} \& {Lazarian}(1998)}]{draine_lazarian98}
{Draine}, B.~T. \& {Lazarian}, A. 1998, \apj, 508, 157

\bibitem[{{Draine} \& {Li}(2001)}]{draine_li01}
{Draine}, B.~T. \& {Li}, A. 2001, \apj, 551, 807

\bibitem[{{Draine} \& {Li}(2007)}]{draine_li07}
---. 2007, \apj, 657, 810

\bibitem[{{Lazarian} \& {Draine}(1997)}]{lazarian&draine_97}
{Lazarian}, A. \& {Draine}, B.~T. 1997, \apj, 487, 248

\bibitem[{{Leger}(1988)}]{leger_88}
{Leger}, A. 1988, {in Polarized Radiation of Circumstellar Origin, ed. G. V.
  Coyne, A. F. J. Moffat, S. Tapia, A. M. Magalhaes, R. E. Schulte-Ladbeck \&
  N. C. Wickramasinghe (Vatican Observatory), 769-795}

\bibitem[{{Leger} \& {Puget}(1984)}]{leger&puget_84}
{Leger}, A. \& {Puget}, J.~L. 1984, \aap, 137, L5

\bibitem[{{Purcell}(1979)}]{purcell_79}
{Purcell}, E.~M. 1979, \apj, 231, 404

\bibitem[{{Rouan} {et~al.}(1992){Rouan}, {Leger}, {Omont}, \&
  {Giard}}]{rouan_92}
{Rouan}, D., {Leger}, A., {Omont}, A., \& {Giard}, M. 1992, \aap, 253, 498

\bibitem[{{Sellgren} {et~al.}(1988){Sellgren}, {Rouan}, \&
  {Leger}}]{sellgren_88}
{Sellgren}, K., {Rouan}, D., \& {Leger}, A. 1988, \aap, 196, 252

\bibitem[{{Smith} {et~al.}(2007){Smith}, {Draine}, {Dale}, {Moustakas},
  {Kennicutt}, {Helou}, {Armus}, {Roussel}, {Sheth}, {Bendo}, {Buckalew},
  {Calzetti}, {Engelbracht}, {Gordon}, {Hollenbach}, {Li}, {Malhotra},
  {Murphy}, \& {Walter}}]{Smith+Draine+Dale+etal_2007}
{Smith}, J.~D.~T., {Draine}, B.~T., {Dale}, D.~A., {Moustakas}, J.,
  {Kennicutt}, R.~C., {Helou}, G., {Armus}, L., {Roussel}, H., {Sheth}, K.,
  {Bendo}, G.~J., {Buckalew}, B., {Calzetti}, D., {Engelbracht}, C.~W.,
  {Gordon}, K.~D., {Hollenbach}, D.~J., {Li}, A., {Malhotra}, S., {Murphy},
  E.~J., \& {Walter}, F. 2007, \apj, 656, 770

\bibitem[{{Tielens} {et~al.}(1993){Tielens}, {Meixner}, {van der Werf},
  {Bregman}, {Tauber}, {Stutzki}, \&
  {Rank}}]{Tielens+Meixner+vanderWerf+etal_1993}
{Tielens}, A.~G.~G.~M., {Meixner}, M.~M., {van der Werf}, P.~P., {Bregman}, J.,
  {Tauber}, J.~A., {Stutzki}, J., \& {Rank}, D. 1993, Science, 262, 86

\end{thebibliography}

\appendix
\section{Polarization from PAH{\scriptsize s} in brownian rotation}\label{app}
We have assumed above that the internal alignment between the
principal axis $\ba$ and angular momentum $\bJ$ of emitting PAHs can
be parametrized via the probability distribution in \eq{align} both before UV absorption (with $\gamma_0$) and during IR emission (with $\gamma_r\leq\gamma_0$). 
However, if the interval between two subsequent photon absorptions is
longer than the time required for the PAH molecule to collide with its own
mass of gas, and if the IVRET process is ineffective at exchanging energy
between the grain rotation and the (generally colder) vibrational modes,
the molecule will be driven towards ``Brownian rotation''
with internal alignment temperature $T_0\approx T_{\rm gas}$ 
prior to UV absorption.  In
this case, the probability distribution $\ud P_{\epsilon}$ in
\eq{align2} should be used instead of \eq{align} to describe the
grain internal alignment before UV absorption, whereas $\ud P_{\gamma_r}$ in
\eq{align} with internal alignment coefficient
$\gamma_r$ may still be appropriate during IR emission.  If the
molecule angular momentum is randomly oriented in space, the polarized
emission along a direction $\hat{\b{w}}$ (= either $\hat{\b{u}}$ or
$\hat{\b{v}}$ -- see \fig{geom}\pan{a}) for in-plane and
out-of-plane modes from a population of PAHs illuminated by a point-like source is
\begin{equation}
I_{\star, w}^{\parallel,\perp}(\alpha,\epsilon,\gamma_r)\propto\int_{-1}^{1}\ud\cos\theta\int_0^{2\pi}\ud\varphi\int_0^\pi\ud P_{\epsilon}(\beta_0)\int_0^\pi\ud P_{\gamma_r}(\beta_r)\; \bar A_{\star}(\theta,\beta_0)\,\bar F_w^{\parallel,\perp}(\theta,\varphi,\beta_r,\alpha)~~~.
 \end{equation}
The corresponding expression for PAH molecules above the center of a uniform-brightness disk galaxy is
\begin{equation}
I_{\mathrm{gal}, w}^{\parallel,\perp}(\alpha,\epsilon,\gamma_r,\omega)\propto\int_{-1}^{1}\ud\cos\theta\int_0^{2\pi}\ud\varphi\int_0^\pi\ud P_{\epsilon}(\beta_0)\int_0^\pi\ud P_{\gamma_r}(\beta_r)\int_{\cos\omega}^{1}\frac{\ud\cos\theta'}{\Omega} \;\bar A_{\rm{gal}}(\theta,\beta_0,\theta')\,\bar F_w^{\parallel,\perp}(\theta,\varphi,\beta_r,\alpha)~~~,
 \end{equation}
where $\Omega=2\pi(1-\cos\omega)$ is the solid angle subtended by the galactic disk as seen from the emitting molecules. 
The degree of polarization can then be computed from \eq{pol_def}. Let
us define \be
q(\epsilon)\equiv3\,\epsilon\,[1-\sqrt{\epsilon-1}\arcsin(1/\sqrt{\epsilon})]-1
~~~,
\ee
where $\epsilon\equiv I_1/(I_1-I_2)$. Making use of
$h(\gamma)$ as defined in \eq{h_def}, the  polarization for
in-plane and out-of-plane modes in case of uni-directional
illumination is
\begin{eqnarray}
p^\parallel_\star(\alpha,\epsilon,\gamma_r)&=&\frac{3\, \sin^2\alpha}{320\,h(\gamma_r)/q(\epsilon)+3\, \cos^2\alpha-1}~~~,\\
p^\perp_\star(\alpha,\epsilon,\gamma_r)&=&\frac{-3\, \sin^2\alpha}{160\,h(\gamma_r)/q(\epsilon)-3\, \cos^2\alpha+1}~~~,
\end{eqnarray}
whereas for diffuse illumination by a uniform-brightness disk galaxy we have
\begin{eqnarray}
p^\parallel_{\rm gal}(\alpha,\epsilon,\gamma_r,\omega)&=&\frac{3\, \sin^2\alpha}{640\frac{h(\gamma_r)/q(\epsilon)}{(\cos^2\omega+\cos\omega)}+3\, \cos^2\alpha-1}~~~,\\
p^\perp_{\rm gal}(\alpha,\epsilon,\gamma_r,\omega)&=&\frac{-3\, \sin^2\alpha}{320\frac{h(\gamma_r)/q(\epsilon)}{(\cos^2\omega+\cos\omega)}-3\, \cos^2\alpha+1}~~~.
\end{eqnarray}
Comparison of these formulae with
eqs.~(\ref{eq:star1})-(\ref{eq:star2}) and
eqs.~(\ref{eq:galaxy1})-(\ref{eq:galaxy2}) shows that the polarization
results obtained here can be exactly reproduced by the probability
distribution in \eq{align} if we adopt an ``effective'' internal
alignment coefficient $\gamma_0$ before UV absorption satisfying
$2\,h(\gamma_0)=1/q(\epsilon)$.  For $\epsilon=2$, as appropriate for
our model PAH with $I_1/I_2=2$, the solution is $\gamma_0\approx1.0$.
\end{document}